\documentclass[sigconf]{acmart}
\AtBeginDocument{%
  \providecommand\BibTeX{{%
    \normalfont B\kern-0.5em{\scshape i\kern-0.25em b}\kern-0.8em\TeX}}}
\acmConference[]{}{}{}

\usepackage{algorithmic}
\usepackage{graphicx}
\usepackage{textcomp}
\usepackage{xcolor}
\usepackage{xspace}
\usepackage{amsmath}
\usepackage{paralist}
\usepackage[english]{babel} 
\usepackage[utf8]{inputenc}
\usepackage[T1]{fontenc}
\usepackage{zlmtt} 
\usepackage{fancyvrb}
\usepackage{hyperref}
\usepackage{listings}
\usepackage{cleveref}
\usepackage{varwidth}
\usepackage{newverbs}
\usepackage{url}
\usepackage{multirow}
\usepackage{hyperref}

\def\BibTeX{{\rm B\kern-.05em{\sc i\kern-.025em b}\kern-.08em
    T\kern-.1667em\lower.7ex\hbox{E}\kern-.125emX}}

\newcommand{\ResourceType}{\text{RType}\xspace}
\newcommand{\RLC}{\text{RLC}\xspace}
\newcommand{\CCRLC}{\text{RLC\#}\xspace}

\newcounter{example}
\newenvironment{example}[1][]{\refstepcounter{example}\par\medskip
   \noindent \textbf{Example~\theexample. #1} \rmfamily}{$\square$\medskip}

\definecolor{cverbbg}{gray}{0.93}
\definecolor{cverbbgl}{gray}{0.7}

\newenvironment{lcverbatim}
 {\SaveVerbatim{cverb}}
 {\endSaveVerbatim
  \flushleft\fboxrule=0pt\fboxsep=.5em
  \colorbox{cverbbg}{%
    \makebox[\dimexpr\linewidth-2\fboxsep][l]{\BUseVerbatim{cverb}}%
  }
  \endflushleft
}
\newenvironment{llcverbatim}
 {\SaveVerbatim{cverb}}
 {\endSaveVerbatim
  \flushleft\fboxrule=0pt\fboxsep=.5em
  \colorbox{cverbbgl}{%
    \makebox[\dimexpr\linewidth-2\fboxsep][l]{\BUseVerbatim{cverb}}%
  }
  \endflushleft
}

\newverbcommand{\cverb}
  {\setbox\verbbox\hbox\bgroup}
  {\egroup\colorbox{cverbbg}{\box\verbbox}}

\begin{document}

\title{Resource Leak Checker (\CCRLC) for C\# Code using CodeQL}

\author{Pritam Gharat}
\email{t-prgharat@microsoft.com}
\affiliation{%
  \institution{Microsoft Research}
  \country{India}
}
\author{Narges Shadab}
\email{narges.shadab@email.ucr.edu}
\affiliation{%
  \institution{University of California, Riverside}
  \country{USA}
}
\author{Shrey Tiwari}
\email{shreymt@gmail.com}
\affiliation{%
  \institution{Microsoft Research}
  \country{India}
}
\author{Shuvendu Lahiri}
\email{shuvendu.lahiri@microsoft.com}
\affiliation{%
  \institution{Microsoft Research}
  \country{USA}
}
\author{Akash Lal}
\email{akashl@microsoft.com}
\affiliation{%
  \institution{Microsoft Research}
  \country{India}
}

\begin{abstract}

Resource leaks occur when a program fails to release a finite resource after it is no longer needed. 
These leaks are a significant cause of real-world crashes and performance issues. 
Given their critical impact on software performance and security, detecting and preventing resource leaks is a crucial problem.

Recent research has proposed a specify-and-check approach to prevent resource leaks. 
In this approach, programmers write \emph{resource management specifications} that guide how resources are stored, 
passed around, and released within an application.

We have developed a tool called \CCRLC, for detecting resource leaks in C\# code. 
Inspired by the Resource Leak Checker (\RLC) from the \emph{Checker Framework}, \CCRLC employs \emph{CodeQL}
for intraprocedural data flow analysis. The tool operates in a modular fashion and 
relies on resource management specifications integrated at method boundaries for interprocedural analysis.

In practice, \CCRLC has successfully identified 24 resource leaks in open-source projects and internal proprietary 
Azure microservices. Its implementation is declarative, and it scales well. 
While it incurs a reasonable false positive rate, the burden on developers is minimal, involving the 
addition of specifications to the source code.
\end{abstract}
\label{dummy-label-for-etags}

\maketitle


\section{Introduction}
\label{sec:intro}

A resource leak occurs when a finite resource managed by developers is acquired (via allocation) but is not disposed when it is 
no longer needed.
In managed languages like C\#, these resources can include file handles, database connections, and network connections. 
Resource leaks can lead to resource starvation, performance slowdowns, crashes, and even denial-of-service attacks. 
Despite its critical impact on software performance and security, preventing and detecting resource leaks 
remains a challenging problem.

The motivation of this work is to detect resource leaks in Azure
microservices written in C\# code.
To achieve this, we have developed a resource leak checker called \CCRLC. 
Inspired by the Resource Leak Checker (\RLC)~\cite{kellogg2021lightweight} —an open-source tool for Java—our \CCRLC targets C\# code and leverages 
\emph{CodeQL} for data flow analysis. 

Like its Java counterpart, \CCRLC focuses on intraprocedural analysis, ensuring that methods like `Close' or `Dispose', 
used for releasing or disposing of resources, are called appropriately on relevant objects (references to resources). 
Violations of this \emph{must-call} property are reported as resource leaks. 

\RLC, the Java counter-part, performs three intraprocedural analyses to verify compliance with this property 
(details available in \cref{sec:background}).
A key insight in \RLC is that checking the must-call property is an \emph{accumulation} problem.
This verification doesn't necessitate heavyweight whole-program interprocedural alias analysis. 
Although the accumulation analysis remains sound even in the absence of alias information, 
it becomes imprecise without such information. To address this limitation, \RLC uses local aliasing and a 
resource management specification language. Developers write these specifications detailing how resources are 
managed within an application—how they are stored, passed around, and ultimately released. These specifications, 
integrated at method boundaries, facilitate interprocedural data flow analysis, rendering the accumulation analysis modular.

\CCRLC, our resource leak checker, leverages the data flow analysis module of CodeQL to verify the must-call property. 
Unlike the accumulation problem tackled by its predecessor, the resource leak checking in \CCRLC can be framed as a 
\emph{graph-reachability} problem.
CodeQL constructs a data flow graph that captures the flow of data between its nodes. These nodes correspond to 
relevant expressions within a program. For instance:
for a resource leak checker, a relevant expression may include a call to a constructor
that allocates a resource or a call to a method \texttt{Dispose}. For a null-dereference check,
a relevant expression may include a NULL assignment or a pointer dereference.

\CCRLC harnesses the expressive power of code-pattern matching in CodeQL. By identifying all relevant expressions, 
it ensures comprehensive coverage for resource leak detection.
CodeQL provides predicates (akin to APIs) that allow \CCRLC to verify the must-call property. It checks for the existence of 
one or more data flow paths between relevant nodes.

\CCRLC uses the local data flow of CodeQL and 
to address unsoundness (because of calls), \CCRLC uses specifications that 
capture interprocedural data flow. 
Since Java and C\# are similar in syntax with both being statically typed object-oriented languages, \CCRLC readily  
borrows the specification language from \RLC.

\Cref{sec:background} provides a brief background of CodeQL and \RLC. \Cref{sec:design,sec:implementation} describe the design 
and implementation details of \CCRLC. It also makes a comparison between \CCRLC and \RLC.
The evaluation of our tool \CCRLC to detect resource leaks on open-source projects and Azure microservices
is described in \cref{sec:evaluation}
and \cref{sec:conclusions} concludes the paper.

\section{Background}
\label{sec:background}

This section gives a brief background on CodeQL used by \CCRLC,
intraprocedural analyses of \RLC, and the resource management specification
language.

\subsection{An Overview of CodeQL}

CodeQL~\cite{codeql} is an analysis engine primarily developed to automate security checks.
CodeQL supports extensive set of languages and frameworks that are widely used\footnote{https://codeql.github.com/docs/codeql-overview/supported-languages-and-frameworks/}.
CodeQL is available for platforms like Linux, Windows and macOS. 
GitHub provides the CodeQL command-line interface and CodeQL for Visual Studio Code for performing CodeQL analysis on open source codebases.

CodeQL treats code like data and provides an expressive code-pattern-matching mechanism.
CodeQL extracts a single relational representation of each source file in order to create a database. 
A CodeQL query (written in an object-oriented query language called QL) is used to query the database
to extract useful information from the source code.
A simple query that finds a redundant \texttt{if} statement in the source code is given below.
\begin{llcverbatim}
from IfStmt ifstmt, BlockStmt block 
where ifstmt.getThen() = block and 
      block.isEmpty()
select ifstmt, "if-statement is redundant." 
\end{llcverbatim}
The \texttt{from} clause in the above query declares variables, the \texttt{where} clause represents the logical formula and 
the \texttt{select} clause specifies the results to display for the variables that meet the conditions defined in the \texttt{where} clause.

A CodeQL user can model security vulnerabilities, bugs, and other errors as a query which is run against a CodeQL database
to find potential bugs that are highlighted directly in the source files.
Many CodeQL queries implement data flow analysis to find such vulnerabilities and bugs.

CodeQL provides a \texttt{DataFlow} module 
in which the libraries implement data flow analysis on a program by modeling its data flow graph. 
A data flow graph models data flow through the program at runtime. Nodes in the data flow graph 
represent semantic elements (such as expressions, parameters) that carry values at run time, and
edges in the graph represent the data flow between the program elements. 
The nodes and edges in a data flow graph are governed by the analysis (the data flow to be tracked);
for null dereference check, a node in a data flow graph could be an assignment to NULL or a pointer dereference.

CodeQL allows a user to specify what needs to be computed without worrying about how it is computed.
Thus, a user only has to declaratively specify the nodes and edges of a data flow graph for the analysis.
CodeQL constructs the data flow graph by using classes modeling the program elements that represent the graph's nodes. 
The flow of data between the nodes is modeled using predicates (similar to APIs) to compute the graph's edges.
The construction of a precise data flow graph poses several challenges such as source code not being available, precise alias information and scalability for large real-world programs.

To overcome this problem, CodeQL provides two kinds of data flow.
CodeQL can capture local data flow within a method as well as global data flow across methods. 
The local data flow graph contains nodes that represent expressions of a single method. The calls within a method 
are ignored and hence the information computed by the local data flow is under-approximated (unsound). On the other hand,
the global data flow graph includes nodes representing expressions of different methods. 
The global data flow captures the local data flow as well as the additional data flow across the methods through calls. 
The information computed by global data flow is over-approximated.
CodeQL's documentation claims that the local data flow is generally fast, efficient and precise as compared to the global
data flow. 

In \CCRLC, we use the expressive power of code-pattern-matching of CodeQL to specify the nodes in the data flow graph for resource leak analysis.
\CCRLC uses the local data flow for performing the intraprocedural analysis.

\subsection{Accumulation-based Resource Leak Checker \RLC}
\label{sec:rlc-desc}

We first describe the three intraprocedural analyses that are used to verify the must-call property in \RLC~\cite{kellogg2021lightweight}.
Later, we describe the specification language that helps to make \RLC modular.

\subsubsection{Verifying Must-Call Property}

\RLC defines a \emph{must-call} property that ensures that methods such as \texttt{Close} or \texttt{Dispose}
used for releasing/disposing resources are \emph{must}-called on a relevant object (a reference to a resource).
Violation of the must-call property is reported as a resource leak. \RLC performs three intraprocedural analyses
to verify the must-call property.
\begin{enumerate}
	\item An analysis that computes an over-approximated set of methods (referred to as \emph{MustCall} set) 
		that must be called on a relevant object.
	\item An analysis that computes an under-approximated set of methods (referred to as \emph{CalledMethods} set)
		that are actually called on a relevant object.
	\item An analysis that compares the two sets and checks if the MustCall methods are invoked 
		by a relevant object before it goes out of scope. 
\end{enumerate}
For an acquired resource \texttt{r}, 
a set \texttt{Ref} is computed at every program point such that \texttt{r} can be accessed using any reference in \texttt{Ref}.
This set is an under-approximation of all the references that can be used to access a given resource in a program. 
The references in \texttt{Ref} are either must aliases or resource
aliases (i.e., different objects that all hold the same underlying
resource); any reference in \texttt{Ref} has the ability 
to satisfy the must-call property for the resource \texttt{r}.
The must-call property for an acquired resource \texttt{r} guarantees that the methods (for releasing the resource) are called
before all the references to \texttt{r} become unreachable.
The verification of must-call property at a program point \texttt{s} is performed as follows:
\begin{align*}
\exists \texttt{e} \in \texttt{Ref}\;\; \texttt{MustCall(e)}\; \subseteq\; \texttt{CalledMethods(e)}
\end{align*}

A resource \texttt{r} is leaked if there exists a program point \texttt{s} where the must-call property for \texttt{r} is violated 
and the set of relevant objects to \texttt{r} (\texttt{Ref}) at \texttt{s} is empty. 
This indicates that the resource \texttt{r} may not be released before all the relevant objects to \texttt{r} become unreachable
along some path in the program.

\begin{figure}[t]
\small
\begin{tabular}{c@{\;\;}|c}
\begin{lstlisting}
1:a = new Socket(...);
  if(...)
2:  ...
  else {
3:  a.Close();
4: }
\end{lstlisting}
&
\begin{tabular}{@{\,}c@{\,}|@{\,}c@{\,}|@{\,}c@{\,}}
\hline
Stmt. & \texttt{Ref} & \begin{tabular}{c} Must-call \\ satisfiability \end{tabular}
\\ \hline \hline
1 & \{\texttt{a}\} & No
\\ \hline
2 & \{\texttt{a}\} & No
\\ \hline
3 & \{\texttt{a}\} & Yes
\\ \hline
4  & \{\} & No
\\ \hline
\end{tabular}
\end{tabular}
\caption{An example demonstrating the data flow analysis performed by the resource leak checker. The must-call property is an all paths property. At the join point of the two branches after the else statement, the must-call property is unsatisfiable along the path $1-2-4$ but satisfiable along the path $1-3-4$, hence it is unsatisfiable at 4.}
\label{fig:must-call-eg}
\end{figure}

\begin{example}
In \Cref{fig:must-call-eg},
a resource is acquired in line 1.
The set \texttt{MustCall} for every relevant object of \texttt{r} contains a single method \texttt{Close}.
The set \texttt{CalledMethods(a)} is empty at line 1 whereas at
line 3, the set \texttt{CalledMethods(a)} contains \{\texttt{Close}\}.

The must-call property is an all paths property. At the join point of the else statement, 
the must-call property is unsatisfiable along the path $1-2-4$ 
but satisfiable along the path $1-3-4$, hence it is unsatisfiable at 4.
A leak is reported at statement 4 because the must-call property is not satisfied and the set of relevant objects to 
\texttt{Socket} allocated in statement 1 is empty.
\end{example}

\subsubsection{Resource Management Specifications}

A developer communicates with the resource leak checker by writing specifications describing
how the resources are stored or passed around in an application and how these resources are released. 
The specifications added to the method boundaries capture the interprocedural data flow (without looking at the method bodies) thereby enabling the accumulation analysis to be modular.
These specifications also capture the must-alias information across method boundaries.

The specifications are expressed as attributes enclosed within
square brackets ([...]) in C\# whereas in Java, they are expressed as annotations which start with
an at-sign (@). In this paper, we use the C\# style of specifications. 

{\bf Attribute \texttt{MustCall(m)} for type.} The attribute \texttt{MustCall(m)} on a type definition indicates that the method \texttt{m} must be called on every instance of the type.
In C\#\footnote{In Java all types that implement the interface \texttt{java.io.Closeable} must define the method \texttt{close} and the annotation 
\texttt{MustCall(close)} on the type indicates
that every instance of the type must call the method \texttt{close} that is used for explicitly releasing the unmanaged resources.},
all the types that implement the interface \texttt{System.IDisposable} must define the method \texttt{Dispose} and the attribute \texttt{MustCall(Dispose)} on the type
indicates that every instance of the type must call the method \texttt{Dispose} that is used for explicitly releasing the unmanaged resources.
\begin{example}
In the example, the user-defined type \texttt{Container} implements the interface \texttt{System.IDisposable}. 
\begin{lcverbatim}
1. [MustCall(Dispose)]
2. class Container() : IDisposable {
3.   ...
14.  public void Dispose() {
15.    // release unmanaged resources
16.  }
17.  ...
18.  public static void Main() {
19.     Container c =  new Container();
20.     ...
23.     c.Dispose();
24.  }
25.}
\end{lcverbatim}
	The resource leak checker verifies the must-call property by checking that every instance of type \texttt{Container} (\texttt{c} on line 19)
	must call method \texttt{Dispose}. This is specified by \texttt{MustCall(Dispose)} associated to the type \texttt{Container}.
\label{eg:MustCallAttribute}
\end{example}

{\bf Attribute \texttt{Owning} for parameter and return-type of methods.} The attribute \texttt{Owning} expresses 
ownership and ownership transfer.
When two references access the same resource, the attribute indicates which one of the two is responsible for satisfying
the must-call property.

The \texttt{Owning} attribute associated to a return-type of a method \texttt{m} indicates that a resource is allocated 
and returned by \texttt{m}. 
The attribute represents the transfer of ownership of deallocation of the resource from the method \texttt{m} to its callers.

When the attribute \texttt{Owning} is associated to a parameter \texttt{p}, it indicates that the parameter \texttt{p} has the obligation of deallocating the resource.
There is a transfer of ownership from the caller to the callee of the method.
\begin{example}
In the example, a resource created at line 3 in method \texttt{createSocket} is returned. 
The call to \texttt{createSocket} (line 5 inside method \texttt{perform}) 
represents the resource being returned by the method and a reference \texttt{so} created at the call-site has the obligation of releasing the resource.
The resource is then passed as an argument to the method \texttt{closeSocket}. Inside method \texttt{closeSocket}, the resource is released by calling
method \texttt{Dispose} on line 18. 
\begin{lcverbatim}
1   [Owning]
2:  Socket createSocket() {
3:    return new Socket(...); 
4:  }
5:  void perform() {
6:    Socket so = createSocket();
7:    ...
15:   closeSocket(so);
16: }
17: void closeSocket([Owning] Socket s) {
18:   s.Dispose();
19: } 
\end{lcverbatim}
There is a transfer of ownership of disposing the resource from the callee (\texttt{createSocket}) to the caller (\texttt{perform}) by returning the
	resource. This is represented by the attribute \texttt{Owning} associated to the return-type of method \texttt{createSocket}. 
	There is another transfer of ownership of disposing the resource from the caller (as an argument 
	\texttt{so} in \texttt{perform}) to the callee (through a formal parameter \texttt{s} of the method \texttt{closeSocket}). This is represented by associating
	the \texttt{Owning} attribute to the parameter \texttt{s}.
\label{eg:OwningParameter}
\end{example}

{\bf Attribute \texttt{Owning} for field/property of a class.} When a resource is stored in a field/property \texttt{f} of a class \texttt{C}, the verification of the must-call property usually spans multiple methods.
The \texttt{Owning} attribute associated to \texttt{f} indicates that \texttt{f} has the obligation of disposing the resource.
For the verification of the must-call property of \texttt{f}, \RLC performs the following checks:
\begin{enumerate}
	\item The attribute \texttt{MustCall(d)} is associated to the type \texttt{C} for some method \texttt{C.d()}.
	\item The method \texttt{C.d()} must always invoke \texttt{this.f.m()} thereby satisfying the must-call property of \texttt{f}.
\end{enumerate}
\begin{example}
In the example, the user-defined type \texttt{Container} contains a field \texttt{socket} 
that is initialized in the constructor at line 8. 
The type \texttt{Socket} implements the interface \texttt{System.IDisposable} and hence has the attribute \texttt{MustCall(Dispose)} associated to it.
	Thus, every instance of type \texttt{Socket} must invoke method \texttt{Dispose}. Thus, the method \texttt{this.f.m()} above is \texttt{Socket.Dispose()}.
\begin{lcverbatim}
1. [MustCall(Dispose)]
2. class Container() {
3. 
4.  [Owning]
5.  private readonly Socket socket;
6.
7.  public Container() {
8.    socket = new ...
9.  }
10. [EnsuresCalledMethods(socket, Dispose)]
11. public void Dispose() {
12.   socket.Dispose();
13. }
14. public static void Main() {
15.   Container c = new Container();
      ...
25.   c.Dispose();
26.  }
27. }
\end{lcverbatim}

A resource is allocated only when the variable \texttt{c} is instantiated on line 15 through the call to the constructor. The method
\texttt{Container.Dispose()} (method \texttt{C.d()} above) invokes method \texttt{Socket.Dispose()} on the field \texttt{socket}.  

Condition 1 is checked through the \texttt{MustCall(d)} on type \texttt{C}. Condition 2 
requires a postcondition (\texttt{this.f.m} is invoked within the method) on method \texttt{C.d()}. 
	\texttt{EnsuresCalledMethods(f, m)} is used for representing this postcondition on \texttt{m}.
\label{eg:OwningField}
\end{example}

{\bf Attribute \texttt{EnsuresCalledMethods(f, m)} for a method.} The attribute on a method indicates that it is responsible for releasing the resource 
by calling method \texttt{f.m}().

In \Cref{eg:OwningField}, the field \texttt{socket} is disposed by method \texttt{Dispose} (line 12). 
Thus, \texttt{EnsuresCalledMethods(socket, Dispose)} is associated to \texttt{Dispose} where the second parameter to the attribute is a method
defined by the type \texttt{Socket}.
The attribute represents the postcondition that the associated method disposes the resource referenced by the \texttt{Owning} field/property.

{\bf Attribute \texttt{CreateMustCallFor(f)} for a method.} The attributed method allocates a new resource and assigns it to
an \texttt{Owning} field/property \texttt{f}.
\begin{example}
We modify \Cref{eg:OwningField} by making the field \texttt{socket} non-read-only with an addition of a new method \texttt{reset} and  a
modified \texttt{Main}.	
A non-read-only field can be reinitialized (method \texttt{reset}) and hence references a new resource (line 39). 
A call to \texttt{reset} (line 25) allocates a new resource. This is represented by the 
attribute \texttt{CreateMustCallFor(f)} associated to \texttt{reset}.
\begin{lcverbatim}
14. public static void Main() {
15.  Container c = new Container();
     ...
25.  c.reset();
26.  ...
32.  c.Dispose();
33. }
34. [CreateMustCallFor(socket)]
35. public void reset()
36. {
37.   if(socket != NULL)
38.     socket.Dispose();
39.   socket = new ...;
40. }
\end{lcverbatim}
Note that the checker should check if the older resource is disposed before a new resource is allocated; otherwise it reports a leak. 
In the example, the older resource referenced by \texttt{socket} is disposed (line 38) before the reassignment in the method \texttt{reset}. 
\label{eg:CreateMustCallForAttribute}
\end{example}

{\bf Attribute \texttt{MustCallAlias} for parameter and return-type of a method.} The attribute represents \emph{resource-aliasing} relationship;
resource aliases are distinct references that refer to the same resource. 
Thus, the resource can be disposed by calling \texttt{Dispose} on any one of the resource aliases.
\begin{example}
In the example, a resource is allocated on line 8 which is passed as an argument to 
method \texttt{createAlias} on line 17. Inside \texttt{createAlias}, no new resource is created; however
a must-alias to the parameter \texttt{s} is being returned (line 4). Thus, at call site 17, the argument 
	\texttt{sock} to \texttt{createAlias} and the call expression \texttt{createAlias(sock)} refer to the same 
	resource. The reference \texttt{new\_sock} is aliased
	 to \texttt{createAlias(sock)} because of the assignment in line 17. Thus, \texttt{sock} and 
	 \texttt{new\_sock} are aliases and hence the resource can be disposed 
	 by calling \texttt{Dispose} on either \texttt{sock} or \texttt{new\_sock}.
\begin{lcverbatim}
1. [MustCallAlias]
2. public Socket createAlias([MustCallAlias]
3.      Socket s) {
4.    Socket new_s = s;
5.    return new_s;
6. }
7. public static void Main() {
8.    Socket sock = new Socket(...);
9.    ...
17.   Socket new_sock = createAlias(sock);
18.   new_sock.Dispose();
19.}
\end{lcverbatim}
	The aliasing relationship between \texttt{sock} and \texttt{createSocket(sock)} 
	is represented by the attribute \texttt{MustCallAlias}
associated to parameter \texttt{s} (corresponding to the argument \texttt{sock}) and the return-type of 
the \texttt{createAlias}
	(corresponding to \texttt{createAlias(sock)}). 
The two references are called resource aliases because they refer to the same resource acquired
on line 8.
\label{eg:MustCallAlias1}
\end{example}

The attribute \texttt{MustCallAlias} is useful to represent wrapper types. 
An instance of a wrapper type wraps or contains an instance of another type. 
The instance of a wrapper type has a field that stores the instance of another type.
Thus, the instance of the wrapper class and the wrapped instance are resource aliases because they reference the same resource.
\begin{example}
In the example, \texttt{SWrapper} is a wrapper type that wraps an instance of 
	\texttt{Socket}. 
\begin{lcverbatim}
1. [MustCall(Dispose)]
2. public class SWrapper {
3.  [Owning]
4.  private Socket socket;

5.  [MustCallAlias]
6.  public SWrapper([MustCallAlias] Socket s) {
7.     this.socket = s;
8.  }
9.  ...
14. public static void Main() {
15.    Socket sock = new Socket(...);
16.    ...
25.    SWrapper wrap_sock = new SWrapper(sock);
27.    wrap_sock.Dispose();
28. }
29.}
\end{lcverbatim}
	The references \texttt{sock} and \texttt{SWrapper(sock)}
	are resource aliases identified by the attribute \texttt{MustCallAlias} (added to parameter \texttt{s} and the constructor \texttt{SWrapper})
	that refer to the same resource allocated on line 15.
\label{eg:MustCallAlias2}
\end{example}


\section{Design of the CodeQL Query for \CCRLC}
\label{sec:design}

\CCRLC uses the intraprocedural data flow analysis of CodeQL to verify the must-call property.
The interprocedural data flow is captured using the specifications (attributes in C\#)
added to the method boundaries.
The data flow analysis in CodeQL involves construction of a data flow graph. The nodes in the graph represent 
relevant program elements for a particular analysis. For a resource leak checker, calls to constructors, calls to methods
such as \texttt{Dispose} are relevant program elements and form the nodes in the data flow graph. The edges
connecting the nodes represent  a data flow between them.
The graph also contains nodes that represent aliases to the relevant program elements. 

A CodeQL query written for resource leak analysis contains three essential components. The first two components are general and are essential for any data flow analysis.
The last one is specific to resource leak analysis.
\begin{enumerate}
	\item Specification of nodes that represent the source (where the data flow begins) and the sink (where the data flow ends) 
		of a data flow. CodeQL refers to them as \emph{source} and \emph{sink} nodes respectively.
		For a resource leak checker, a source node may represent a call to a constructor that allocates a resource whereas a call to \texttt{Dispose} may form a sink node.

		Note that we are performing intraprocedural data flow analysis and hence the source and sink nodes must belong to the same method.
	\item For every source node, a check for the existence of a sink node and a data flow path between the source and the corresponding sink node.
		The data flow path between a source and a sink node indicates that the resource allocated by the source node is the one being disposed by
		the sink node.
	\item The analysis reports a resource leak if there exists at least one control-flow path along which the acquired resource is not released (including exceptional paths). 
		Thus, simply checking the existence of a sink node for a given source node is not sufficient. The query also needs to 
		check if there exists a sink node along every control-flow path
		from the source node to the exit of the method.
\end{enumerate}

\subsection{Specification of Source and Sink Nodes}
\label{sec:src-sink}

\CCRLC uses the expressive power of code-pattern-matching of CodeQL to provide specifications 
of the source and the sink nodes. 
These nodes could be pattern-matched as calls to methods with appropriate attributes.
This is because \CCRLC being modular (similar to \RLC) only looks at the attributes of a method when a call to the method is encountered and not the method body.

A source node in a data flow graph for a resource leak checker could be pattern-matched as:
\begin{enumerate}
	\item A \texttt{new} expression (call to a constructor) that allocates a resource (e.g. line 3 in \Cref{eg:OwningParameter})\label{itm:srp1}.
\item A call to a method whose return-type has \texttt{Owning} attribute associated to it (e.g. line 6 in \Cref{eg:OwningParameter})\label{itm:srp2}.
	The \texttt{Owning} attribute associated to a return-type of a method indicates that a resource is allocated inside the method and is being returned to the caller.
	Hence, call to such a method forms a source node.
\item A call to a method with \texttt{CreateMustCallFor} attribute (e.g. line 25 in \Cref{eg:CreateMustCallForAttribute})\label{itm:srp3}. 
	The attribute \texttt{CreateMustCallFor} represents allocation of a new resource within a method. Hence a call to method
		with this attribute is considered as a source node\label{itm:srp4}.
\item A parameter with an \texttt{Owning} attribute (e.g. line 17 in \Cref{eg:OwningParameter})\label{itm:srp5}.
	The \texttt{Owning} attribute associated to a parameter indicates that the parameter is a reference to a resource and it is obligated to release the resource. Hence, the
		parameter with an \texttt{Owning} attribute forms a source node.
\end{enumerate}

Similarly, a sink node in a data flow graph for a resource leak checker could be pattern-matched as:
\begin{enumerate}
	\item A call to a method \texttt{Close} or \texttt{Dispose} (line 23 in \Cref{eg:MustCallAttribute})\label{itm:sip1}.
	\item A return expression of a method whose return-type has an \texttt{Owning} attribute (line 3 in \Cref{eg:OwningParameter})\label{itm:sip2}.
		The return expression returns a reference to a resource to the caller thereby transferring the ownership of releasing the resource to the caller and hence
		this forms the sink node.
	\item A call to a method whose parameter has an \texttt{Owning} attribute (line 15 in \Cref{eg:OwningParameter})\label{itm:sip3}.
	\item A call to a method with \texttt{EnsuresCalledMethods} attribute associated to it (line 25 in \Cref{eg:OwningField})\label{itm:sip4}.
\end{enumerate}

The specifications of source and sink nodes are general and cover most of the patterns that capture how the resources are acquired, passed around, stored, and released in an application.

\subsection{Dataflow between Source and Sink Nodes}

A data flow path between a source and sink node in a data flow graph exists only when the resource acquired at the source node
is released by the sink node. Consider, the following code snippet:
\begin{lcverbatim}
1. Socket s1 = new Socket(...);
2. Socket s2 = new Socket(...);
3. s2.Dispose();
4. s1.Dispose();
\end{lcverbatim}
CodeQL pattern-matches the \texttt{new Socket(...)} expression in lines 1 and 2 as the source nodes in the data flow graph.
The calls to \texttt{Dispose} on lines 3 and 4 are pattern-matched as the sink nodes. The source node in line 1 is aliased to the reference \texttt{s1} and the one in line 2
is aliased to \texttt{s2}. 
There is a data flow path between the source node in line 1 and the sink node in line 4 (not line 3) as the data flows through the alias \texttt{s1} which is a receiver object for the call
to \texttt{Dispose} in line 4.
Similarly, there is a data flow path between the source node in line 2 and the sink node in line 3.

CodeQL can capture local data flow within a method as well as global data flow across methods. 
The local data flow graph contains nodes that represent expressions within a single method. The calls to other methods
are ignored and hence the information computed by the local data flow is under-approximated (unsound). 
Predicate \texttt{localFlow(node1, node2)} holds if there exists a data flow path from 
\texttt{node1} to \texttt{node2} where both \texttt{node1} and \texttt{node2} belong to the same method.

On the other hand,
the global data flow graph includes nodes representing expressions of different methods. 
The global data flow captures the local data flow as well as the additional data flow across the methods through calls. 
The information computed by global data flow is over-approximated.
Predicate \texttt{hasFlow(node1, node2)} holds if there exists a data flow path from 
\texttt{node1} to \texttt{node2} where \texttt{node1} and \texttt{node2} may belong to different methods.

CodeQL claims that the local data flow is generally fast, efficient and precise as compared to the global
data flow. 
\CCRLC uses the local data flow of CodeQL and addresses the unsoundness issue by using attributes 
to capture the interprocedural data flow.

\subsection{Verification of Must-Call Property}

The existence of a data flow path between a source and a sink node is a necessary but not a sufficient condition for guaranteeing
no resource leak. The existence of a data flow path only guarantees that the resource acquired at a source node is released by the sink node
along some path. However, a resource is said to be leaked if there exists at least one control-flow path along which the resource is not released (sink does not exist).
\begin{example}
In the example, socket is allocated in line 2 (source node) and disposed in line 14 (sink node). 
However, if an exception occurs before line 14, then
the socket will not be disposed and a leak will occur. 
In this case, although there exists a sink node and a data flow path exists between the source and the sink node, 
there is still a leak.
This is because there exists a control-flow path (exceptional path) along which the sink node does not exist.
\begin{lcverbatim}
1. try {
2.   Socket s = new Socket(...);
3.   ...
14.   s.Dispose();
15. }
16. catch(...) {...}
\end{lcverbatim}
If we modify the above code snippet by adding a call \texttt{s.Dispose()} in the catch-block, then there will be no resource leak. 
In this case, there will be two sink nodes for a source
node in line 2. Both the nodes will have data flow paths with the source node and all the control flow paths from the source node to the end of the method (in this case code snippet)
contain a sink node.
\label{eg:exception}
\end{example}

\section{Implementation in CodeQL}
\label{sec:implementation}

In this section, we describe the details of the CodeQL query (\CCRLC) that conforms to the design (\Cref{sec:design}).
Later, we make a comparison between \CCRLC and \RLC. We then share our experience of adopting CodeQL to develop \CCRLC.

\subsection{Modelling Source and Sink Nodes}

We track the resources using the types of the references. The set \ResourceType represents such types.
The pattern-matching of source and sink node includes checking if the type of the node is a member of \ResourceType. 
A type \texttt{t} belongs to \ResourceType if either
\begin{itemize}
	\item \texttt{t} implements the \texttt{System.IDisposable} interface, or
\item \texttt{t} has \texttt{Owning} attribute associated to one or more fields, or
\item \texttt{t} has a \texttt{MustCall} attribute associated to its definition, or
\item \texttt{t} is a \texttt{CollectionType} (like Vectors, Arrays) where the type of the elements is in \ResourceType.
\item \texttt{t} is a subtype of another type in \ResourceType.
\end{itemize}

We present the snippets of CodeQL query for pattern-matching source node (refer to the list in \Cref{sec:src-sink}).

{\bf A \texttt{new} expression (call to a constructor):} CodeQL pattern-matches the \texttt{new} expression 
as an \texttt{ObjectCreation}. 
The query pattern-matches any \texttt{o} (representing a \texttt{new} expression) as a source node
if the type of \texttt{o} matches that of one of the types in \ResourceType.
\begin{llcverbatim}
exists(ObjectCreation o | o.getType() in RType)
\end{llcverbatim}
Some of the examples of the source node with this pattern can be found in line 19 (\Cref{eg:MustCallAttribute}), 
line 3 (\Cref{eg:OwningParameter}), lines 8 and 15 (\Cref{eg:OwningField}).

{\bf A call to a method whose return-type has \texttt{Owning} attribute:} 
The query pattern-matches the source node as a \texttt{call} where the callee \texttt{c}
has an attribute \texttt{a} of type \texttt{Owning}.
\begin{llcverbatim}
exists(Call call, Callable c, Attribute a |
   c = call.getARuntimeTarget() and
   a = c.getAnAttribute() and
   a.getType().hasName("Owning"))
\end{llcverbatim}
The CodeQL predicate \texttt{getARuntimeTarget()} returns a set of methods
that are called at a call site. CodeQL computes an over-approximation of the call graph that also accounts for virtual calls.

An example of the source node with this pattern can be found in line 6 (\Cref{eg:OwningParameter}).

{\bf A call to a method with \texttt{CreateMustCallFor} attribute:} 
The query pattern-matches the source node as a \texttt{call} where the callee \texttt{c}
has an attribute \texttt{a} of type \texttt{CreateMustCallFor}.
\begin{llcverbatim}
exists(Call call, Callable c, Attribute a |
   c = call.getARuntimeTarget() and
   a = c.getAnAttribute() and
   a.getType().hasName("CreateMustCallFor"))
\end{llcverbatim}
An example of the source node with this pattern can be found in line 25 (\Cref{eg:CreateMustCallForAttribute}).

{\bf A parameter with an \texttt{Owning} attribute:} 
A parameter \texttt{p} with \texttt{Owning} attribute is pattern-matched as:
\begin{llcverbatim}
exists(Parameter p, Attribute a |
   a = p.getAnAttribute() and
   a.getType().hasName("Owning"))
\end{llcverbatim}
An example of the source node with this pattern can be found in line 17 (\Cref{eg:OwningParameter}).

We present the snippets of CodeQL query for pattern-matching sink node (refer to the list in \Cref{sec:src-sink}).

{\bf A call to a method \texttt{Close} or \texttt{Dispose}:} 
The sink node is pattern-matched to any \texttt{call} where the callee \texttt{c} is either \texttt{Close}
or \texttt{Dispose}.
\begin{llcverbatim}
exists(Callable c, Call call |
   c = call.getARuntimeTarget() and
   (c.hasName("Close") or c.hasName("Dispose")))
\end{llcverbatim}
An example of the sink node with this pattern can be found in line 23 (\Cref{eg:MustCallAttribute}).

{\bf A return expression of a method whose return-type has an \texttt{Owning} attribute:}
The sink node is pattern-matched as a return expression \texttt{e} of method \texttt{m}
that has an \texttt{Owning} attribute associated to its return-type. 
\begin{llcverbatim}
exists(Callable m, Expr e, Attribute a |
   m.canReturn(e) and a = m.getAnAttribute() and
   a.getType().hasName("Owning"))
\end{llcverbatim}
The CodeQL predicate \texttt{canReturn} holds if the expression \texttt{e} is the return expression of method \texttt{m}.

An example of the sink node with this pattern can be found in line 3 (\Cref{eg:OwningParameter}).

{\bf A call to a method whose parameter has an \texttt{Owning} attribute:}
The sink node is a \texttt{call} to a method \texttt{c} whose one of the parameter \texttt{p} 
has an \texttt{Owning}
attribute associated to it.
\begin{llcverbatim}
exists(Call call, Parameter p, 
       Callable c, Attribute a |
   c = call.getARuntimeTarget() and
   p = c.getParameter(_) and
   a = p.getAnAttribute() and
   a.getType().hasName("Owning"))
\end{llcverbatim}
The don't care ``\_'' in the query above indicates any parameter of \texttt{c} that is further restrained to have an
\texttt{Owning} attribute.

An example of the sink node with this pattern can be found in line 15 (\Cref{eg:OwningParameter}).

{\bf A call to a method with \texttt{EnsuresCalledMethods} attribute:}
The sink is a \texttt{call} to a method \texttt{c} that has  
\texttt{EnsuresCalledMethods} associated to it.
\begin{llcverbatim}
exists(Call call, Callable c, Attribute a |
   c = call.getARuntimeTarget() and
   a = c.getAnAttribute() and
   a.getType().hasName("EnsuresCalledMethods"))
\end{llcverbatim}

An example of the sink node with this pattern can be found in line 25 (\Cref{eg:OwningField}).

\subsection{Modelling Dataflow between Source and Sink Nodes}

CodeQL predicate \texttt{localFlow} checks for local data flow
(where source and sink nodes belong to the same method) between the two nodes. 
The data flow across methods is captured through the specifications/attributes added at the method boundaries.
These specifications also provide information about resource aliases which are not captured by \texttt{localFlow}.

Recall that resource aliases are distinct expressions that reference the same resource. 
The predicate \texttt{localFlow} cannot capture this data flow because the aliasing happened in the callee of the method. 
The attribute \texttt{MustCallAlias} captures this aliasing relationship. 
We introduce \texttt{isResourceAlias} predicate to identify resource aliases.

The predicate \texttt{isResourceAlias} checks for a call to a method
whose parameter and return-type has \texttt{MustCallAlias} attribute.
Predicates \texttt{isMustCallAliasPar} and \texttt{isMustCallAliasMethod} are defined
to check if the parameter and the return-type of the method is associated with \texttt{MustCallAlias} attribute.

\begin{llcverbatim}
predicate isResourceAlias(node1, node2) {
exists(AssignableDefinition def, Call call, 
       Callable c, Parameter p, Expr arg |
   call = def.getSource() and 
   c = call.getARuntimeTarget() and 
   p = c.getParameter(_) and
   arg = call.getArgumentForParameter(p)
   and node2.asExpr() = def.getSource() and 
   node1.asExpr() = arg and 
   isMustCallAliasMethod(c) and 
   isMustCallAliasPar(p))
}
\end{llcverbatim}
An assignment \texttt{def} contains a call in its RHS (CodeQL predicate \texttt{def.getSource()} gives the RHS of \texttt{def}). The argument to the call \texttt{arg} (\texttt{node1}) and the expression representing
the \texttt{call} (\texttt{node2}) are resource aliases. 
This is represented by a \texttt{MustCallAlias} attribute associated to the parameter \texttt{p} of the callee \texttt{c} (corresponding 
to \texttt{arg}) and the return-type of \texttt{c}.

\begin{example}
In the constructor \texttt{Wrapper}, the parameter \texttt{s} and
field \texttt{socket} are must aliases. This aliasing relationship
is represented by the \texttt{MustCallAlias} attribute for 
the parameter and the return-type (implicit in case of constructor) of the constructor \texttt{Wrapper}.

\begin{lcverbatim}
1. class Wrapper {
2.  [Owning]
3.  private Socket socket;
4.  [MustCallAlias] 
5.   public Wrapper([MustCallAlias] Socket s) {
6.     socket = s;
7.   }
8. 
9.   static void perform() {
10.    Socket s = new Socket(...);
11.    Wrapper w = new Wrapper(s); 
12.    w.Dispose(); // same as s.Dispose();
13.  }
14.}
\end{lcverbatim}
A call to the method \texttt{Wrapper} on line 11 indicates 
that the argument being passed to the constructor (\texttt{s} in this case) and 
the return-variable (represented by the call \texttt{Wrapper(s)}) are resource aliases.
The expressions \texttt{Wrapper(s)} and \texttt{s} refer to the same resource. 
\label{eg:resource-alias}
\end{example}

We define the predicate \texttt{isAlias} that captures all the data flow between two nodes
\texttt{node1} and \texttt{node2}.
\begin{llcverbatim}
predicate isAlias(node1, node2) {
   DataFlow::localFlow(node1, node2) or
   isResourceAlias(node1, node2) or
   exists(DataFlow::Node n | 
        isAlias(node1, n) and isAlias(n, node2))
}
\end{llcverbatim}
Nodes \texttt{node1} and \texttt{node2} are either
local aliases (\texttt{localFlow})
or resource aliases (\texttt{isResourceAlias}) 
or a combination of the two.
In \Cref{eg:resource-alias}, 
\texttt{Wrapper(s)} and \texttt{s} are resource aliases. The references \texttt{w} and \texttt{Wrapper(s)} are aliases
captured by \texttt{localFlow}. 
Thus, \texttt{w} and \texttt{s} are aliases because of the transitive relationship 
(last condition of the \texttt{isAlias} predicate).

\subsection{Modelling the Verification of Must-Call Property}

For a given source node \texttt{src}, verifying the existence of a sink node and a data flow path
between \texttt{src} and the sink node is necessary but not sufficient condition.
A resource should be released along all control-flow paths from the source node to the exit of the method.

The predicate \texttt{notDisposed} checks if a resource corresponding to a source node (first parameter)
is released along all control-flow paths. The second argument is a control-flow node that keeps changing as we 
traverse the control-flow graph.
We begin from the exit of a method and traverse the control-flow graph (CFG) backwards 
(CodeQL predicate \texttt{getAPredecessor} is used for backward traversal of CFG).
We stop the traversal of a control-flow path
either when we encounter a sink node or a source node. If we encounter a source node, it means that there exists a control-flow
path from the source node to the exit of the method which does not contain sink. If we encounter a sink node, then we stop exploring
the control-flow path indicating that the resource is released along this path.
\begin{llcverbatim}
predicate notDisposed(src, nd) {
   nd = src.getControlFlowNode() or
   notDisposed(src,nd.getAPredecessor()) and not
   exists(DataFlow::Node sink | 
      sink.getControlFlowNode() = nd 
      and isAlias(src, sink))
}
\end{llcverbatim}

Violation of the condition indicates a potential resource leak:
\begin{llcverbatim}
notDisposed(src, 
       src.getEnclosingCallable().getExitPoint())
\end{llcverbatim}
In \Cref{eg:exception}, for the source node in line 2, there exists a control-flow path along which a sink node does not
exist (the exception path from any line 3-13 to the catch-block) and hence a leak is reported. However, if the example
is modified with an addition of a sink node (\texttt{s.Dispose()}) in the catch-block then \texttt{notDisposed} will
not report a leak.

\subsection{Working of \CCRLC}

In this section, we demonstrate how \CCRLC works on our examples.

In \Cref{eg:MustCallAttribute}, the type \texttt{Container} is a resource type because it implements the
\texttt{System.IDisposable} interface. The source is a \texttt{new} expression in line 19 whereas the sink 
is a call to \texttt{Dispose} in line 23. \CCRLC does not report a resource leak in this case because a sink node exists with
a data flow path from source to sink (indicating the resource is released). Also, the sink node exists along all the control-flow
paths from the source to the exit of \texttt{Main}.

In \Cref{eg:OwningParameter}, inside method \texttt{createSocket}, the source node is the expression \texttt{new Socket(...)}
(line 3) whereas the sink node is the return expression (also line 3). Thus, \CCRLC does not report a leak within method
\texttt{createSocket}. In method \texttt{perform}, the source node is call to \texttt{createSocket} in line 6 (call to a method
with \texttt{Owning} attribute associated to the return-type of the method)
and the sink node is call to \texttt{closeSocket} in line 15 (call to a method whose parameter has an \texttt{Owning} attribute).
The data flow between source and sink node is verified by \texttt{localFlow}. 
Hence, \CCRLC reports no resource leak in
\texttt{perform}. In method \texttt{closeSocket}, the source node is the parameter \texttt{s} (line 17) 
because it has the \texttt{Owning}
attribute and the sink node is call to \texttt{Dispose} (line 18). 
In this case, the data flow between source and sink node is verified by \texttt{localFlow}. 
\CCRLC does not report a leak within \texttt{closeSocket}. 
This example demonstrates the modularity of \CCRLC; the analysis is intraprocedural and yet sound because the 
interprocedural data flow is captured through the attributes at the method boundaries.

In \Cref{eg:OwningField}, the source node is a \texttt{new} expression in line 15 and sink node is call to \texttt{Dispose}
in line 25. \CCRLC does not report a leak because of the existence of the data flow path between the two nodes and all control-flow
paths are covered by the sink node from source to exit of \texttt{Main}.

In \Cref{eg:CreateMustCallForAttribute}, there are two source nodes (lines 15 and 25). The source node in line 25
is a call to a method \texttt{reset} that reassigns the \texttt{Owning} field \texttt{socket}. The sink node in line 32
is a call to a method \texttt{Dispose} that has \texttt{EnsuresCalledMethods(socket, Dispose)} indicating that
\texttt{Dispose} releases the resource stored in the field \texttt{socket}.
\CCRLC reports no resource leak in \texttt{Main} because for each source node there exists a sink node (line 32). 
However, inside \texttt{reset}, \CCRLC verifies that the older resource is disposed before the reassignment.
Since the older resource is released (line 38) before the reassignment, \CCRLC does not report a leak within \texttt{reset}.

In \Cref{eg:MustCallAlias1}, there is a source node in \texttt{Main} (line 8). In line 17,
\texttt{sock} and \texttt{createAlias(sock)} are resource aliases because of the \texttt{MustCallAlias} attribute
associated to the parameter and the return-type of \texttt{createAlias}.
The transitive closure of aliases makes \texttt{sock} and \texttt{new\_sock} as aliases 
(identified by \texttt{isAlias}).
Hence the resource (corresponding to the source node in line 8) can be released using \texttt{sock} or \texttt{new\_sock}.
The call to \texttt{Dispose} in line 18 forms the sink node and hence no leak is reported by \CCRLC.

\subsection{Comparison between \CCRLC and \RLC}
\label{sec:csharp-vs-java}

Although, the development of our tool \CCRLC is inspired from \RLC, the two tools differ in terms of 
design choices and implementation.
In this section, we compare and contrast our tool \CCRLC with \RLC. 

\subsubsection{Language-dependent Differences}

\RLC is used for Java code and is developed as a part of Checker Framework for data flow analysis.
\CCRLC is used for C\# code and uses CodeQL framework for
data flow analysis.
Though Java and C\# are similar in syntax with both being statically typed and object-oriented languages, there are
few differences that \CCRLC has to account for.

In C\#, the \texttt{System.IDisposable} interface requires implementation of a method \texttt{Dispose}
that performs all object cleanup.  This is similar to the JDK's \texttt{java.io.Closeable} interface that
requires the implementation of method \texttt{close} for cleanup.

Java has a language construct \texttt{try-with-resources} which is not present in C\#.
The try-with-resources statement ensures that each resource is closed at the end of the statement execution. 
C\# has an analogous language construct, \texttt{using} statement.
The \texttt{using} declaration calls the \texttt{Dispose} method on the instance 
when it goes out of scope. The \texttt{using} statement ensures that \texttt{Dispose}
is called even if an exception occurs within the \texttt{using} block. 
\CCRLC represents a \texttt{using} statement as a sink node because the call to \texttt{Dispose} is performed
internally.

On major difference between Java and C\# 
is the exception handling mechanism.
In practical cases, resource leaks are often detected along exceptional flow paths. Thus, the mechanism of handling 
exceptions is important for detecting resource leaks. 

Java supports the concept of \emph{checked} 
and \emph{unchecked} exceptions.
Checked exceptions, as the name suggests, are checked at compile time. If some code within a method 
throws a checked exception, then the method must either handle the exception or it 
must specify the exception using the throws keyword.
In C\#, all exceptions are unchecked, so it is not forced by the compiler to either handle or 
specify the exception. It is up to the developers to specify or catch the exceptions.

Note that both \RLC and \CCRLC are unsound in handling unchecked exceptional paths.
The below example in C\# illustrates the unsoundness of \CCRLC.
\begin{example} In the example below, statement 2 could potentially throw a \texttt{DivideByZeroException}.
As a result, the control flow may not reach statement 3 after executing statement 2 and hence the resource
allocated in statement 1 may not be released in the event of an exception causing a resource leak.
\begin{lcverbatim}
1. var s = new Socket(...);
2. int z = x/y;
3. s.Dispose();
\end{lcverbatim}

Since this exception is not managed by the developer (no try-catch-finally block or throws specification),
\CCRLC incorrectly asserts that the sink node (line 3) 
is present along all control-flow paths from the source node (line 1) to the exit of the method.
Hence no resource leak is reported.

If this exception was managed by the developer, say for example, all the statements are enclosed within a try-block,
then the predicate \texttt{notDisposed} is satisfied, thereby reporting a resource leak. This is because there exists a 
control-flow path from line 2 in the try-block to the catch-block that does not contain a sink node.
Hence, an external exceptions analysis is required for the unmanaged exceptions for \CCRLC to be sound. 
A separate exceptions analysis for sound resource leak detection is left for future work.
\end{example}

Note that in Java, the above code snippet will not compile because \texttt{DivideByZeroException} is a checked exception.
In Java, all the practical exceptions are checked and hence the need for exceptions analysis in \RLC is not critical as
compared to \CCRLC.

\subsubsection{Differences in Design}

Recall that \RLC views resource leak detection as an accumulation problem (\Cref{sec:rlc-desc}). 
In \CCRLC, the must-call property is verified by checking the existence of a sink node 
and a data flow path between the source 
and the sink node in the data flow graph constructed by CodeQL. 
The must-call property verification in \CCRLC can be viewed as a reachability problem and not an accumulation problem as in case of \RLC.

\subsubsection{Similarities between \CCRLC and \RLC}

\CCRLC is similar to \RLC in a way that it adopts the specification language (from \RLC) for capturing the interprocedural data flow.
The process of writing specifications for the code 
and the tool using these specifications to capture the interprocedural data flow in a modular fashion 
is identical to that of \RLC.


\begin{table}[t]
\begin{tabular}{@{}c@{\;}|@{\,}c@{\,}|@{\,}c@{\,}|c|c|@{\,}c@{\,}}
& LoC & Resources & TP & FP & Attributes 
\\ \hline 
Lucene.Net & 609,754 & 284 & 8 & 12 & 66 
\\ \hline
EFCore & 883,195 & 176 & 0 & 19 & 132 
\\ \hline
Service 1 & 670,988 & 149 & 7 & 2 & 17
\\ \hline
Service 2 & 194,765 & 263 & 6 & 3 & 22 
\\ \hline
Service 3 & 170,471 & 33 & 3 & 1 & 16 
\\ \hline
Total & 25,29,173 & 905 & 24 & 37 & 253
\end{tabular}
\caption{Verifying the absence of resource leaks. ``LoC'' is lines of non-comment, non-blank C\# code. ``Resources''
is the number of resources created by an application. ``TP'' are true positive warnings. ``FP''
are where \CCRLC reported a potential leak, but manual analysis revealed that no leak is possible. ``Attributes'' are
the specifications added to the code.}
\label{tab:result1}
\end{table}

\subsection{Experience with CodeQL}

This section describes our experience of using CodeQL for the design and development of \CCRLC.

\CCRLC leverages only the local data flow of CodeQL that is generally fast, 
efficient and precise as compared to the global data flow analysis (interprocedural analysis). 
\CCRLC relies on the attributes added to the code for interprocedural data flow thereby making the tool modular.

The \texttt{localFlow} predicate of CodeQL is unsound in presence of calls. 
It also does not account for field dereferences. \CCRLC resolves the issue of calls by using attributes added to method boundaries in the code.
For field dereferences, \CCRLC
defines the predicate \texttt{isFieldAlias} that captures the reads and writes through fields (similar to resource aliases). 
For the example below, \texttt{localFlow} fails to register that \texttt{t.f} is aliased to \texttt{s}. As a result, the analysis reports a false-positive for the resource allocated at line 1.
\begin{lcverbatim}
1. Socket s = new Socket(...);
2. t.f = s;
3. t.f.Close();
\end{lcverbatim}
In \texttt{isFieldAlias}, \texttt{node1} and \texttt{node2}
represent field write (\texttt{fw}) and field read (\texttt{fr}) respectively.
\begin{llcverbatim}
predicate isFieldAlias(node1, node2) {
   exists(FieldWrite fw, FieldRead fr, Field f |
   f.getAnAccess() = fw and fw = node1.asExpr()
   and
   f.getAnAccess() = fr and node2.asExpr() = fr)
}
\end{llcverbatim}
Thus, the data flow between the assignments at lines 1 and 3 through field \texttt{t.f} is captured by the predicate \texttt{isFieldAlias}.

The \texttt{isAlias} predicate from earlier is now updated to also accommodate 
the aliasing through field dereferences.
\begin{llcverbatim}
predicate isAlias(node1, node2) {
   DataFlow::localFlow(node1, node2) or
   isResourceAlias(node1, node2) or
   isFieldAlias(node1, node2) or
   exists(DataFlow::Node n | 
      isAlias(node1, n) and isAlias(n, node2))
}
\end{llcverbatim}

CodeQL constructs an over-approximation of the call-graph that also accounts for virtual calls. Thus, there is 
no separate alias analysis required to resolve virtual calls. The over-approximated version of the call-graph
introduces some imprecision but does not lead to the unsoundness in the analysis that uses the call-graph.

CodeQL provides a framework for a developer to provide specification for nodes in the data flow graph without worrying about how the graph is constructed. 
CodeQL also provides predicates to query the graph for checking the data flow paths between specific nodes.
The framework offers a way to the developer to specify what to compute (declaratively) without worrying about how to compute. 
This makes the implementation of an analysis simple. However, in our experience, the debugging of a CodeQL query is time-consuming and complicated (no tool for debugging).

\section{Evaluation} 
\label{sec:evaluation}

We implemented \CCRLC using CodeQL \cite{codeql} version 2.11.4.
In this section, we describe the evaluation of \CCRLC on open-source projects and Azure microservices.
We also compare \CCRLC with an existing (naive) query in the CodeQL repository 
and finally we list the limitations of \CCRLC.

\subsection{Case Studies}

We selected two open-source C\# projects; we used the latest version of the source code that was available when we began.
Lucene.NET is a high performance search library for .NET whereas Entity Framework (EF Core) is an open source object–relational mapping framework for ADO.NET.
In addition, we analyzed three internal proprietary Azure microservices.
We also modelled micro-benchmarks inspired by the study for resource leaks on Azure code. 
These benchmarks include test-cases from the Checker Framework that we were able to translate from Java to C\#. 

For each application, our methodology is as follows.
\begin{inparaenum}
\item We build a CodeQL database for the source code on which our tool \CCRLC works. We run a simple CodeQL query that counts the number of resources allocated
	in each application. This number works as a deciding factor in selecting the applications for the evaluation.
\item We manually annotate the source code for each application with appropriate attributes. We iteratively ran
	\CCRLC to correct our annotations.
\item We manually classified each warning generated by \CCRLC as an actual resource leak (true positive) or as false warning.
\end{inparaenum}

\begin{table}[t]
\begin{tabular}{c|c}
	Attribute & Count
\\ \hline \hline
\texttt{Owning} & 124
\\ \hline 
\texttt{EnsuresCalledMethods} & 59
\\ \hline 
\texttt{MustCallAlias} & 78
\\ \hline 
\texttt{MustCall} & 57 
\\ \hline 
\texttt{CreateMustCallFor} & 19 
\\ \hline 
	Total & 337
\end{tabular}
\caption{The attributes that we added to the source code.}
\label{tab:result2}
\end{table}


\begin{table}[t]
	\begin{tabular}{c|c|c|c|c|}
	& \multicolumn{2}{c|}{\CCRLC} & \multicolumn{2}{c|}{\begin{tabular}{c} Naive \\ CodeQL \\ query \end{tabular}} 
	\\  \cline{2-5}
	& TP & FP & TP & FP 
	\\ \hline
	Lucene.Net & 8 & 12 & 3 & 42 
	\\ \hline
	EFCore & 0 & 19 & 0 & 1  
	\\ \hline
	Service 1 & 7 & 2 & 0 & 38 
	\\ \hline
	Service 2 & 6 & 3 & 0 & 234 
	\\ \hline
	Service 3 & 3 & 1 & 0 & 3 
	\\ \hline
	Total & 24 & 37 & 3 & 318 
\end{tabular}
\caption{Comparison of \CCRLC with an existing (naive) CodeQL query for resource leaks. ``TP'' means True Positives (actual resource leaks) and ``FP'' stands for False Positives.}
\label{tab:result3}
\end{table}

\subsubsection{True and False Positive Examples}
\Cref{tab:result1} summarizes our results. 
\CCRLC found a few resource leaks in four out of five applications. \CCRLC was able to find known resource leaks
in microservices that had caused an incident (high impact) in the past.
The overall precision of \CCRLC on these applications is 39.34\% (24/61).
Annotating source code manually with attributes is a labor-intensive and time-consuming task. To address this, we selectively added attributes only to program elements whose types are library-defined. We chose to ignore warnings generated by the verifier for custom types. Our work~\cite{10.1145/3622858} on inferring resource management specifications also handles custom types. Interestingly, we discovered a greater number of resource leaks by utilizing the inferred attributes in subsequent work. 

We present a few examples of warnings reported by \CCRLC.
\begin{example}
This example of a true positive contains code from Lucene.Net. 
	A \texttt{Stream} (line 86) is wrapped as 
a \texttt{StreamReader} on line 92. The constructor \texttt{UserDictionary} simply uses the argument
\texttt{reader} (of type \texttt{StreamReader}) for reading from a stream. Hence, no attribute is added to the constructor.
\begin{lcverbatim}
84.public void Inform(IResourceLoader loader) {
85. if (uPath != null) {
86.  Stream stream = loader.OpenResource(uPath);
87.  string encoding = userDictionaryEncoding;
88.  if (encoding is null) {
89.   encoding = Encoding.UTF8.WebName;
90.  }
91.  var decode = Encoding.GetEncoding(encoding);
92.  var reader = new StreamReader(stream,decode);
93.  userDictionary = new UserDictionary(reader);
94. }
95. else {
96.  userDictionary = null;
97. }
98.}   
\end{lcverbatim}
Source node is the call to \texttt{OpenResource}.
The call to \texttt{StreamReader} and the argument \texttt{stream} (in the call) are resource aliases because \texttt{StreamReader} is a wrapper class for \texttt{Stream}.
In this example, \CCRLC reports a resource leak in line 86 (no sink node).
A simple fix is to declare \texttt{stream} on line 86 in a ``using'' statement.
\end{example}
\begin{example}
This example of a false positive is a snapshot from one of the Azure microservices.
\begin{lcverbatim}
70.public SqlDataReader executerWithRetry(
71.   SqlCommand command, CommandBehavior cB) {
72. SqlDataReader reader;
73. try {
74.  reader = command.ExecuteReader(cB);
75. }
76. catch (Exception) {
77.  if (command.Connection != null)
78.   command.Connection.Close();
79.  throw;
80. }
81. return reader;
82.}
\end{lcverbatim}
A \texttt{SqlDataReader} is allocated on line 75 inside the try-block provided there is no exception occurring during allocation.
Thus, there exists a path from line 75 to line 81 inside the catch-block along which there exists no sink node. However, along this path, no resource is allocated and hence
this is a case of a false positive. 
Statically identifying such cases is non-trivial.
\label{eg:exception-fp}
\end{example}

The most common cause for false positives (nearly 80\%) is nullness reasoning: some resource is released only if it is non-null but \CCRLC 
expects the resource to be released along every path. \CCRLC handles simple comparisons with \texttt{null}.
A less common cause for false positives (nearly 3\%) is because of conservative handling of exceptions (\Cref{eg:exception-fp}).
The remaining false positives occur when a resource is being allocated and stored in an \texttt{Owning} field (say \texttt{f}) and there is no method defined in the class
that disposes the resource. 
However, when an instance of this class becomes an \texttt{Owning} field (say \texttt{c}) of another class, then the resource is disposed by calling 
\texttt{Dispose} on \texttt{c.f} inside the \texttt{Dispose} method of another class.

\subsubsection{Attributes added to Source Code}

We added one attribute per 7,000 lines of code (\Cref{tab:result2}).
The overhead of adding manual attributes to the source code includes rebuilding the source code and creating a new version of CodeQL database (CodeQL database creation is not incremental).
In order to avoid the repetitive creation of CodeQL database, we add attributes as logical formulae in the query instead of adding them to the code.
An example of one such formula for an \texttt{Owning} parameter \texttt{s} (\Cref{eg:OwningParameter}) is (``\texttt{RLCTests}'' is the namespace):
\begin{llcverbatim}
fileName="RLCTests/SimpleEg.cs" and lineNo="17" 
and elementType="Parameter" and elementName="s" 
and annotation="Owning"
\end{llcverbatim}

\subsection{Comparison to An Existing CodeQL Query}

We compare \CCRLC with an existing CodeQL query (NCQ) \footnote{Naive CodeQL Query available \href{https://github.com/github/codeql/blob/main/csharp/ql/src/API\%20Abuse/NoDisposeCallOnLocalIDisposable.ql}{here.}}. 
NCQ uses the local data flow of CodeQL and does not handle method calls and field dereferences. This makes NCQ unsound because 
the interprocedural data flow is not captured and the aliasing information is incomplete. 
Unlike \CCRLC, NCQ does not have the overhead of adding attributes to the source code.
NCQ is faster than \CCRLC because it completely ignores the method calls.

With an exception of \texttt{Lucene.Net}, NCQ fails to find any resource leaks. 
The number of false positives is very high because the query naively considers every call to constructor as resource allocation.

 
\subsection{Limitations of \CCRLC}

\CCRLC analyzes only the source code that is available. The verification of native code, implementation of unchecked libraries, and code generated dynamically is not performed
by \CCRLC.

\CCRLC is sound with respect to the set \ResourceType that tracks types having \texttt{MustCall} attribute. 
We wrote specifications for the C\# standard library, focusing on IO-related code in the \texttt{System.IO} package.
Any missing specifications of \texttt{MustCall} attribute could lead \CCRLC to miss resource leaks.

If a program statement that could potentially throw an exception is not enclosed within a try-catch-finally block or is not declared to throw an exception (using the \texttt{throws} clause), then \CCRLC misses to track resource leaks on exceptional paths.
Only the developer managed exceptions are handled soundly by \CCRLC.

Although the attribute overhead (one per 7,000 LoC) is small, the need for manually adding attributes to the source code can be viewed as a bottleneck for our approach (specially in legacy code).

Like any practical system, it is possible that there might be errors in the implementation of \CCRLC or in the design of the analysis.
We identify expressions representing resource allocation (source nodes) and resource deallocation (sink nodes) through code-pattern-matching.
We have identified all the common patterns; however there could be corner-cases where we might have missed some nodes. \CCRLC will be unsound if a source node is missed, it will be imprecise (more false positives) if a sink node is missed.
We have mitigated the threat with micro-benchmarks: 88 test cases containing 4,843 lines of non-comment, non-blank code.\footnote{These micro-benchmarks are publicly available \href{https://github.com/microsoft/global-resource-leaks-codeql}{here}}.

\section{Conclusions}
\label{sec:conclusions}

We developed a resource leak checker for C\# code using CodeQL. Inspired by the resource leak checker in the 
Checker Framework, our tool, \CCRLC, shares a modular and lightweight design with its Java counterpart. 
Unlike its predecessor, we approach the resource leak problem as a graph reachability issue, leveraging CodeQL's 
data flow graph construction for analysis. Our experimental evaluation demonstrated the tool's effectiveness, 
identifying 24 resource leaks in both open-source and proprietary C\# code.


\bibliographystyle{ACM-Reference-Format}
\bibliography{references}
\end{document}